%% file: phikgamma.tex
\documentclass[aps,prl,preprint,tightenlines,superscriptaddress,showpacs,byrevtex]{revtex4}
\usepackage{amsmath}
\usepackage{graphicx}
\usepackage{epsfig}
\usepackage{graphicx} 
\usepackage{color}
\usepackage{multirow}

\newcommand{\btopkg}{B \rightarrow \phi K \gamma}
\newcommand{\btopkog}{B^0 \rightarrow \phi K^0 \gamma}
\newcommand{\btopksg}{B^0 \rightarrow \phi K_S^0 \gamma}
\newcommand{\btopkpg}{B^+ \to \phi K^+ \gamma}
\newcommand{\btopkpmg}{B^{\pm} \to \phi K^{\pm} \gamma}

\begin{document}
  \title{\large \bf Observation of radiative {\boldmath $B^0 \rightarrow \phi K^0 \gamma$} decays}
  
  \input{author-conf2009}
  \noaffiliation
  
\begin{abstract}
We report the first observation of radiative decay $\btopkog$ 
using a data sample 
of $772 \times 10^6$ $B\overline{B}$ pairs collected at 
the $\Upsilon(4S)$ resonance 
with the Belle detector at the KEKB asymmetric-energy $e^+e^-$ collider. 
We observe a signal of $35 \pm 8$ events with a 
significance of 
$5.4$ standard deviations
including systematic uncertainties. The measured branching fraction is 
${\cal B}(\btopkog) = (2.66\pm 0.60 \pm 0.32) \times 10^{-6}$.
We also precisely measure
${\cal B}(\btopkpg) = (2.34\pm 0.29 \pm 0.23) \times 10^{-6}$.
The uncertainties are statistical and systematic, respectively.
The observed $M_{\phi K}$ mass spectrum differs
significantly from that expected in a three-body phase-space decay.
\end{abstract}


\maketitle

\par Rare radiative decays of $B$ mesons play an important role in the 
search for physics 
beyond the standard model (SM) of electroweak interactions.
These flavor changing neutral current decays are
forbidden at tree level in the SM, but allowed through
electroweak loop processes. 
The loop can be mediated by non-SM particles
(for example, charged Higgs or SUSY particles), which could 
affect either the branching fraction or the time-dependent $CP$ asymmetry.
\par The current measured inclusive world average branching fraction for
$B \to X_s \gamma$
($(3.55 \pm 0.26) \times 10^{-4}$~\cite{hfag}),
is one standard deviation ($\sigma$)
higher than the SM prediction 
at next-to-next-to-leading order (NNLO) 
$(3.15\pm0.23) \times 10^{-4}$~\cite{theory_misiak}, and still
allows significant new physics contributions to radiative $B$ decays.
Exclusive $b \to s \gamma$ decays have 
also been extensively measured,
but their sum so far accounts only for $44\%$ of the inclusive rate.
Therefore, further measurements of branching fractions for exclusive 
$\btopkg$ modes will improve our understanding of the $b \to s \gamma$ 
process. 
The neutral mode $\btopkog$~\cite{conj} can be used 
to study time-dependent 
$CP$ asymmetry, which is suppressed in the SM by the quark mass 
ratio ($2m_s/m_b$)~\cite{ags1,ags2}.
In several models beyond SM, the photon acquires an appreciable 
right-handed component due to the exchange of a virtual heavy 
fermion in the loop process, 
resulting in large values of time-dependent $CP$ asymmetries.
Due to the narrow width of the $\phi$ resonance, the decay 
$\btopkg$ is well 
separated from the background and can be effectively used 
for measurements of photon momentum over a wide interval.
In addition, this mode can also be used to search for 
a possible contribution from 
kaonic resonances decaying to $\phi K$. 
Furthermore, we can probe the photon polarization 
using the angular distributions of 
the final state hadrons~\cite{pol1,pol2}.
\par
\begin{figure}[htbp]
\begin{center}
\includegraphics[width=9cm]{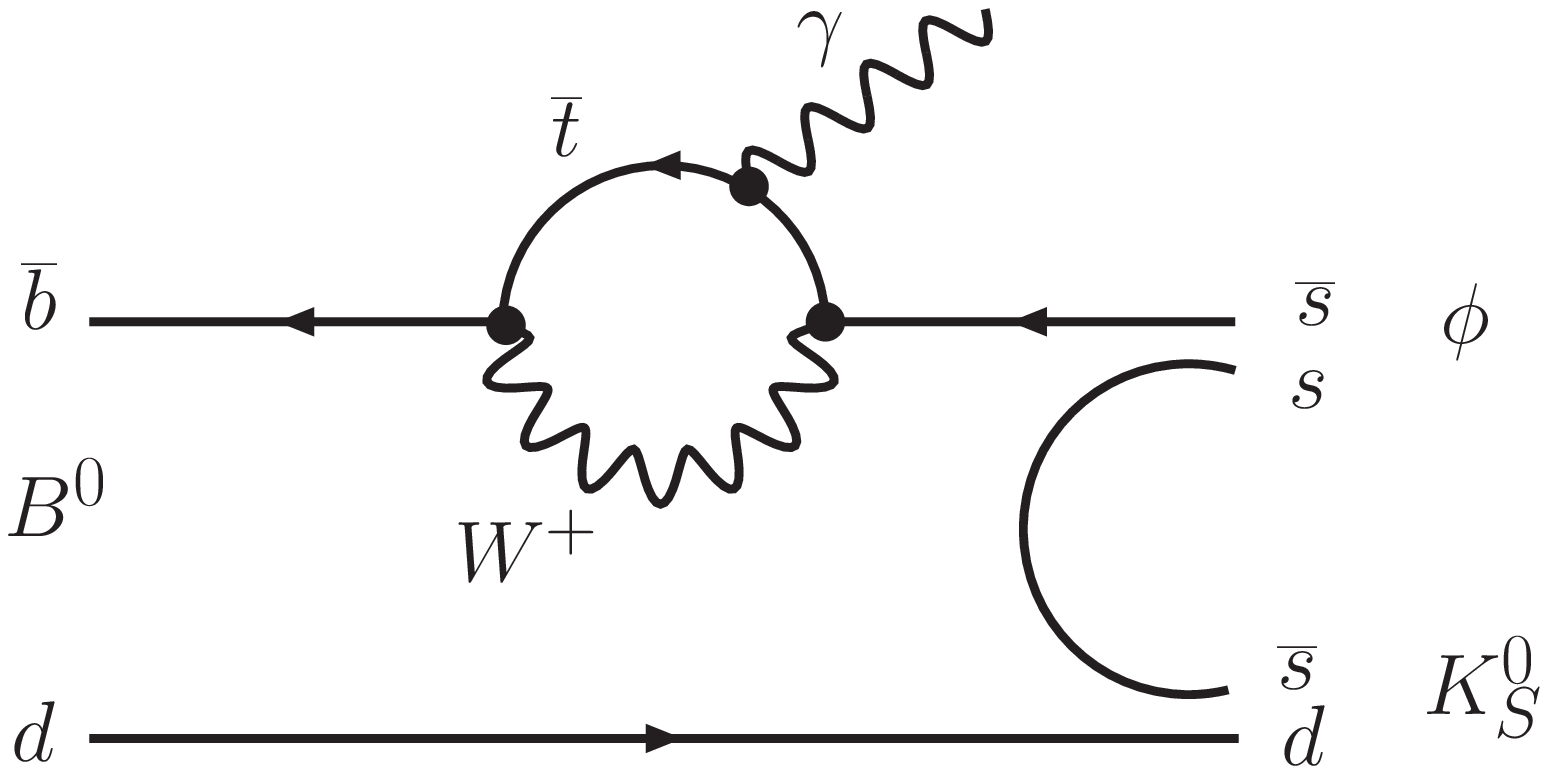}
\end{center}
\caption{Feynman diagram for the radiative penguin penguin 
decay $\btopksg$ with $s\overline{s}$ pair creation.}
\label{fig:penguin}
\end{figure}

 The decay $\btopksg$ can be described by the conventional radiative
penguin diagram with the creation of an additional $s\overline{s}$ pair 
as shown in Fig.~\ref{fig:penguin}.
\par The branching fractions for $\btopkg$ decays have 
already been reported 
by the Belle and BaBar collaborations. Belle measured 
$\mathcal{B}(\btopkpg) = (3.4 \pm 0.9 \pm 0.4) \times 10^{-6}$ and 
$\mathcal{B}(\btopkog) < 8.3 \times 10^{-6}$ 
at the $90\%$ confidence level (C.L.)
using $96 \times 10^6$ $B\overline{B}$ 
pairs~\cite{alex_prl}. BaBar measured 
$\mathcal{B}(\btopkpg) = (3.5 \pm 0.6 \pm 0.4) \times 10^{-6}$ and 
$\mathcal{B}(\btopkog) < 2.7 \times 10^{-6}$ 
at the $90\%$ C.L. using $228 \times 10^6$ $B\overline{B}$ 
pairs~\cite{phikgamma_babar}.
BaBar also reported  the direct $CP$ asymmetry for $\btopkpmg$, 
$\mathcal{A}_{CP} = (-26 \pm 14 \pm 5) \%$.
We report herein the first observation of radiative decay 
$\btopkog$ and an improved measurement of $\btopkpg$ using 
a data sample of $772 \times 10^6$ $B\overline{B}$ pairs
collected at the $\Upsilon(4S)$ 
resonance with the Belle 
detector at the KEKB asymmetric-energy $e^+e^-$ collider~\cite{kekb}.
This data sample is
nearly eight times larger than the sample used in 
our previous measurement~\cite{alex_prl}.
\par The Belle detector is a large-solid-angle magnetic
spectrometer that consists of a silicon vertex detector (SVD),
a 50-layer central drift chamber (CDC), an array of
aerogel threshold Cherenkov counters (ACC),
a barrel-like arrangement of time-of-flight
scintillation counters (TOF), and an electromagnetic calorimeter (ECL)
comprised of CsI(Tl) crystals located inside
a superconducting solenoid coil that provides a 1.5~T
magnetic field.  An iron flux-return located outside 
the coil is instrumented to detect $K_L^0$ mesons and to identify
muons (KLM).  The detector is described in detail elsewhere~\cite{Belle}.
Two different inner detector configurations were used. 
For the first sample of $152 \times 10^6$ $B\overline{B}$ pairs, 
a 2.0 cm radius beampipe
and a 3-layer silicon vertex detector (SVD-I) were used;
for the latter $620 \times 10^6$ $B\overline{B}$ pairs,
a 1.5 cm radius beampipe, a 4-layer silicon vertex detector (SVD-II),
and a small-cell inner drift chamber were used.
A GEANT-based simulation of the Belle detector is used
to produce signal Monte Carlo (MC)~\cite{evtgen} event samples.
\par The signal is reconstructed in the decays
$\btopkpg$ and $\btopksg$, with 
$\phi \to K^+K^-$ and $K_S^0 \to \pi^+\pi^-$.
All the charged tracks used in the reconstruction 
(except for charged pions from $K_S^0$'s)
are required to satisfy a requirement on the distance of closest approach 
to the interaction point (IP) along the 
beam direction, $\left|dz\right|<5$ cm, and 
in the transverse direction, $dr<2$ cm. 
This eliminates poorly reconstructed tracks or tracks that do not
come from the interaction region. 
Charged kaons are identified using a likelihood ratio 
$\mathcal{L} (K/\pi) > 0.6$,
based on information 
from the ACC, TOF and CDC (dE/dx) detectors.
This requirement has an efficiency of $90\%$ for kaons 
with a $8\%$ pion fake rate.
A less restrictive likelihood ratio requirement 
$\mathcal{L} (K/\pi) > 0.4$ 
is applied to the kaon candidates, which are used to 
reconstruct the $\phi$ meson.
The invariant mass of the $\phi$ candidates is required to be within
$-0.01 \;{\rm GeV}/c^2 < M_{K^+K^-}-m_{\phi} < +0.01 \;{\rm GeV}/c^2$,
where $m_{\phi}$ denotes the world-average $\phi$ mass~\cite{pdg}.
\par Neutral kaon ($K_S^0$) candidates are formed from 
the $\pi^+\pi^-$ combinations with 
invariant mass in the range
0.482 GeV/$c^2 < M_{\pi^+\pi^-} < 0.514$ GeV/$c^2$. 
The selected candidates must pass a set of 
momentum-dependent requirements on
impact parameter, vertex displacement, mismatch in the $z$ direction,
and the direction of the pion pair momentum as described 
in the Ref.~\cite{belle_b2s}.
\par The primary signature of this decay 
is a high energy prompt photon.
These are selected from isolated ECL clusters 
within the barrel region
($32^{\circ} < \theta_{\gamma} < 129^{\circ}$,
where $\theta_{\gamma}$ is the polar angle of the 
photon in the laboratory frame)
and center-of-mass system (cms) energy ($E_{\gamma}^{\rm cms}$) 
in the range $1.4$ to $3.4$ GeV.
The selected photon candidates are required to be consistent with 
isolated electromagnetic showers, i.e., $95\%$ of the 
energy in an array of
$5 \times 5$ CsI(Tl) crystals should be concentrated in 
an array of $3 \times 3$ crystals and 
should have no charged tracks associated with it.
We also suppress the background photons
from $\pi^0$($\eta$) $\to \gamma \gamma$ 
using a likelihood $\mathcal{L}_{\pi^0}$($\mathcal{L}_{\eta}$) $< 0.25$, 
calculated for each photon pair consisting of the candidate photon 
and any other photon in the event~\cite{pi0etaveto}.
\par We combine a $\phi$ meson candidate, a charged or neutral 
kaon candidate and the radiative photon to form a $B$ meson. 
The $B$ candidates are identified using 
two kinematic variables: the energy difference 
$\Delta E \equiv E_B^{\rm cms} - E_{\rm beam}^{\rm cms}$ and the
beam-energy-constrained mass 
$M_{\rm bc} \equiv \sqrt{(E_{\rm beam}^{\rm cms})^2 - (p_B^{\rm cms})^2}$,
where $E_{\rm beam}^{\rm cms}$ is the beam energy in the cms, and 
$E_B^{\rm cms}$ and $p_B^{\rm cms}$ are the cms energy and momentum, 
respectively, of the reconstructed $B$ candidate.
In the $M_{\rm bc}$ calculation, the photon momentum is replaced by
$(E_{\rm beam}^{\rm cms} - E_{\phi K}^{\rm cms})$ to improve resolution.
The events that satisfy the requirements
$M_{\rm bc} > 5.2 \;{\rm GeV/}c^2$ and 
$\left|\Delta E\right| < 0.3 \;\rm{GeV}$ 
(defined as the fit region)
are selected for further analysis.
Using MC simulations, we find 
nearly $12\%$ ($3\%$) of events have more than one $B$ candidate
for the $\btopkpg$ ($\btopkog$) mode. 
In case of multiple candidates, we
choose the best candidate based on a series of 
selection criteria, 
which depend on a $\chi^2$ variable using 
the candidate's $\phi$ mass
(and the $K_S^0$ mass in the neutral mode)
as well as
the highest $E_{\gamma}^{\rm cms}$ and  
the highest $\mathcal{L} (K/\pi)$ in the charged mode.
For events with multiple candidates, this selection method chooses the
correct $B$ candidate for the $\btopkpg$ ($\btopkog$) mode
$57\%$ ($69\%$) of the time.
We define the signal region as 
$5.27 \;{\rm GeV/}c^2 < M_{\rm bc} < 5.29 \;{\rm GeV/}c^2$ and 
$-0.08 \;{\rm GeV} < \Delta E < 0.05 \;\rm{GeV}$.
The $\Delta E$ signal region is asymmetric in order to 
include the tail in the lower region due to photon energy 
leakage in the ECL.
\par The dominant background comes from $e^+e^- \to q\overline{q}$
($q = u, d, s,$~or~$c$) continuum events. 
We use two event-shape variables to distinguish the spherically 
symmetric $B\overline{B}$ events from the jet-like continuum events.
A Fisher discriminant~\cite{fisher} is formed from $16$ 
modified Fox-Wolfram moments~\cite{fox} 
and the scalar sum of the transverse momenta.
The second variable is the cosine of the angle between the
$B$ flight direction and the beam axis ($\cos \theta_B$) in the cms frame.
For each variable, we obtain the corresponding signal and background 
probability density functions (PDFs) from large MC samples. 
A likelihood ratio 
$\mathcal{R}_{s/b} = \mathcal{L}_s/(\mathcal{L}_s+\mathcal{L}_b)$
is formed, where $\mathcal{L}_s$ ($\mathcal{L}_b$) denotes the product of 
Fisher discriminant and $\cos \theta_B$ PDFs for the signal (background).
The selection criteria on $\mathcal{R}_{s/b}$ are determined 
by maximizing the figure of merit, $N_S/\sqrt{N_S+N_B}$, 
where $N_S$ ($N_B$) is the expected number of signal 
(continuum) events in the signal region.
We require $\mathcal{R}_{s/b}>0.65$, which removes $91\%$ 
of the continuum while retaining $76\%$ of the signal.
\par In addition to the dominant continuum background, 
various $B\overline{B}$
background sources are also studied. 
In the $\btopksg$ mode,
some backgrounds from $b \to c$ decays, such as
$D^0\pi^0$, $D^0\eta$ and $D^-\rho^+$
peak in the $M_{\rm bc}$ distribution. 
We remove the dominant peaking backgrounds by applying a veto 
to $\phi K_S^0$ combinations consistent with the 
nominal $D$ mass~\cite{pdg}.
Some of the charmless backgrounds, where the $B$ meson decays to
$\phi K^{*}(892)$, $\phi K \pi^0$ and $\phi K \eta$ 
also peak in $M_{\rm bc}$. In these charmless modes, 
one of the photons from a $\pi^0$ or $\eta$ 
may not be detected in
the calorimeter while the other is reconstructed as the signal 
high-energy photon. Therefore, these backgrounds shift towards 
lower $\Delta E$. Another significant background is
non-resonant $B \to K^+ K^- K \gamma$, which peaks in the
$\Delta E$-$M_{\rm bc}$ signal region. 
The fraction of such events is estimated to be
$(12.5\pm6.7)\%$ using the $\phi$ mass sideband, 
$1.05 \;{\rm GeV}/c^2 < M_{K^+K^-} < 1.3 \;{\rm GeV}/c^2$,
in data.
\par
\begin{figure}[htbp]
\begin{center}
\includegraphics[width=8.1cm]{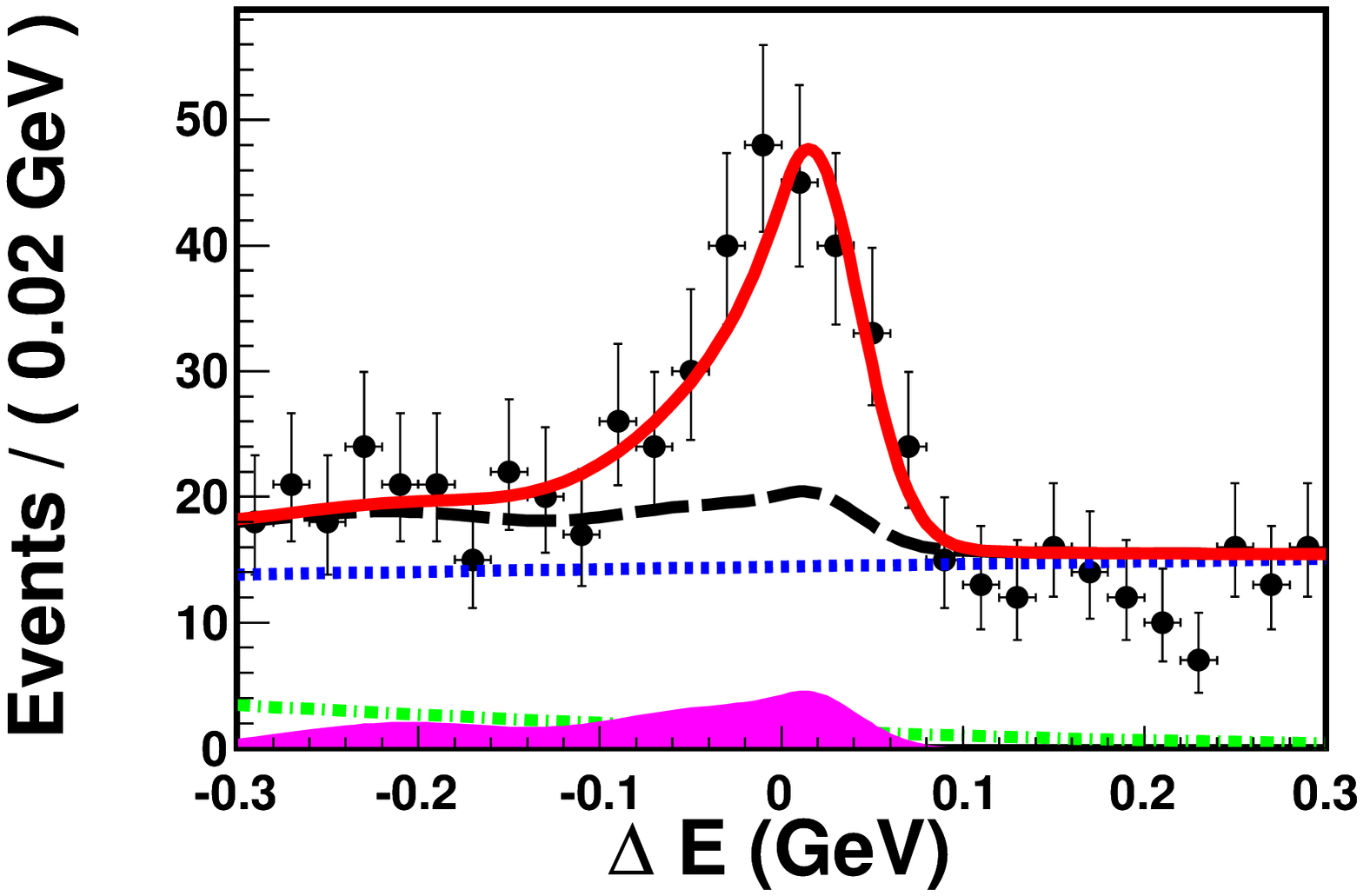}
\includegraphics[width=8.1cm]{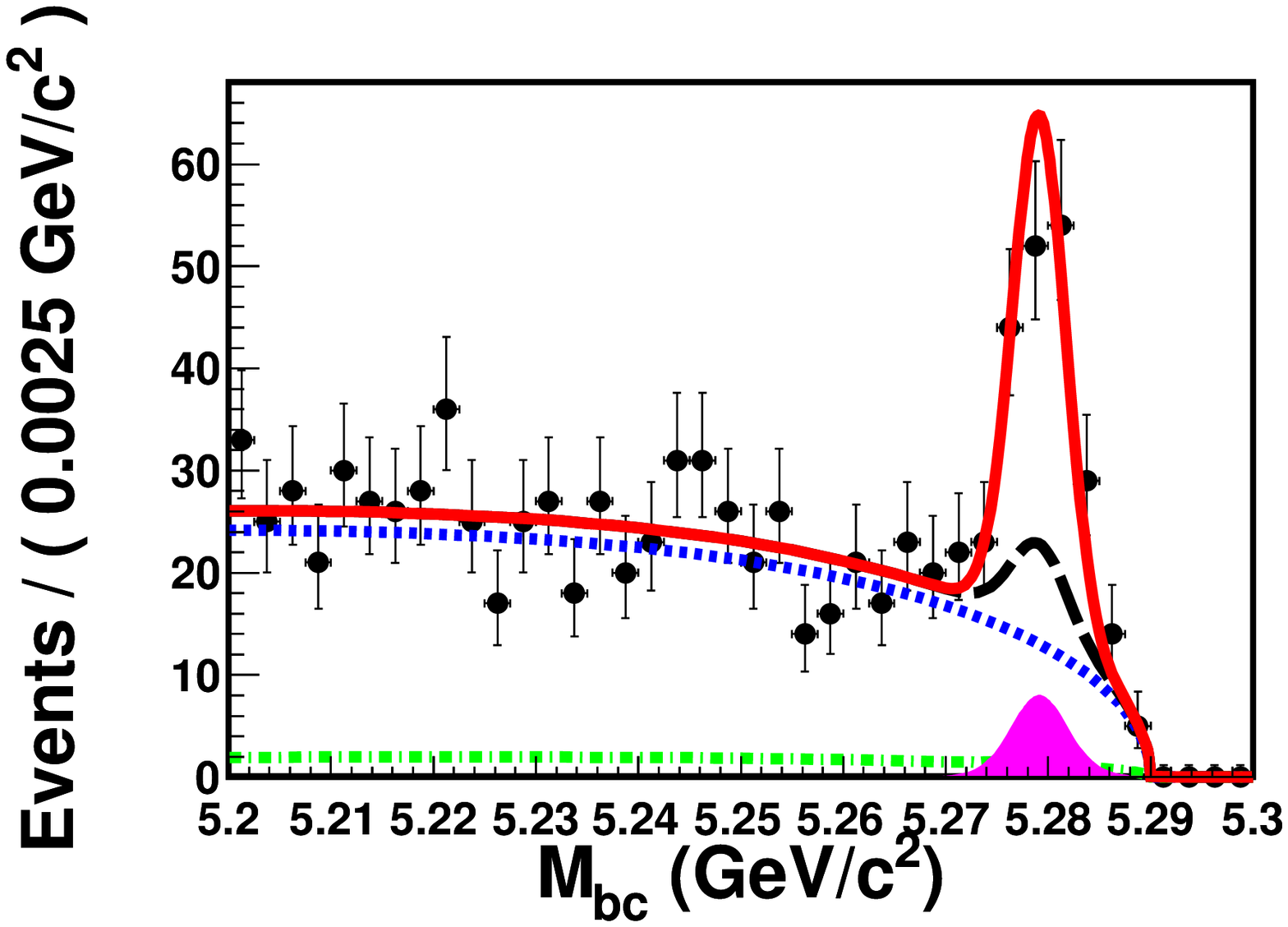}
\includegraphics[width=8.1cm]{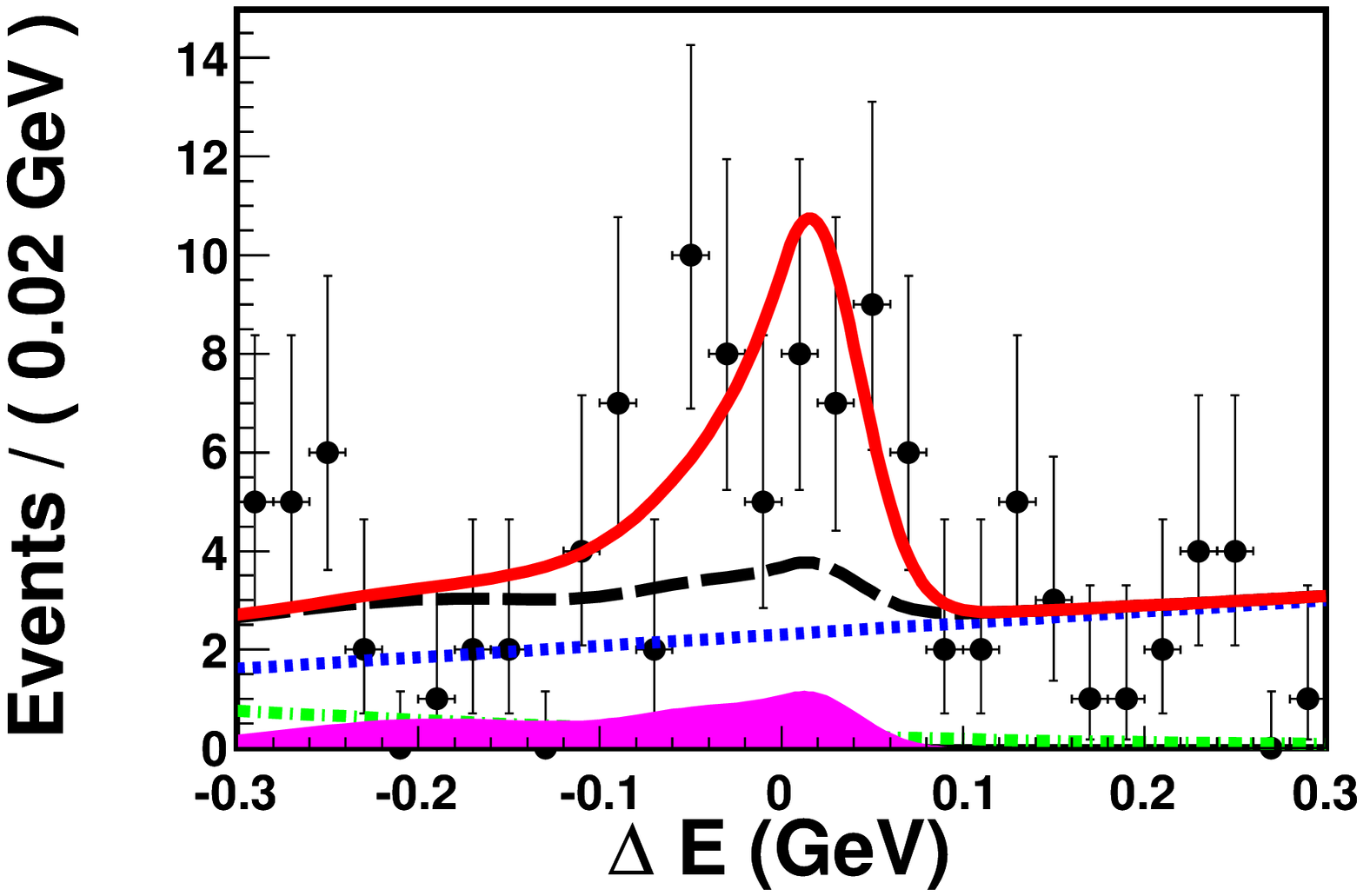}
\includegraphics[width=8.1cm]{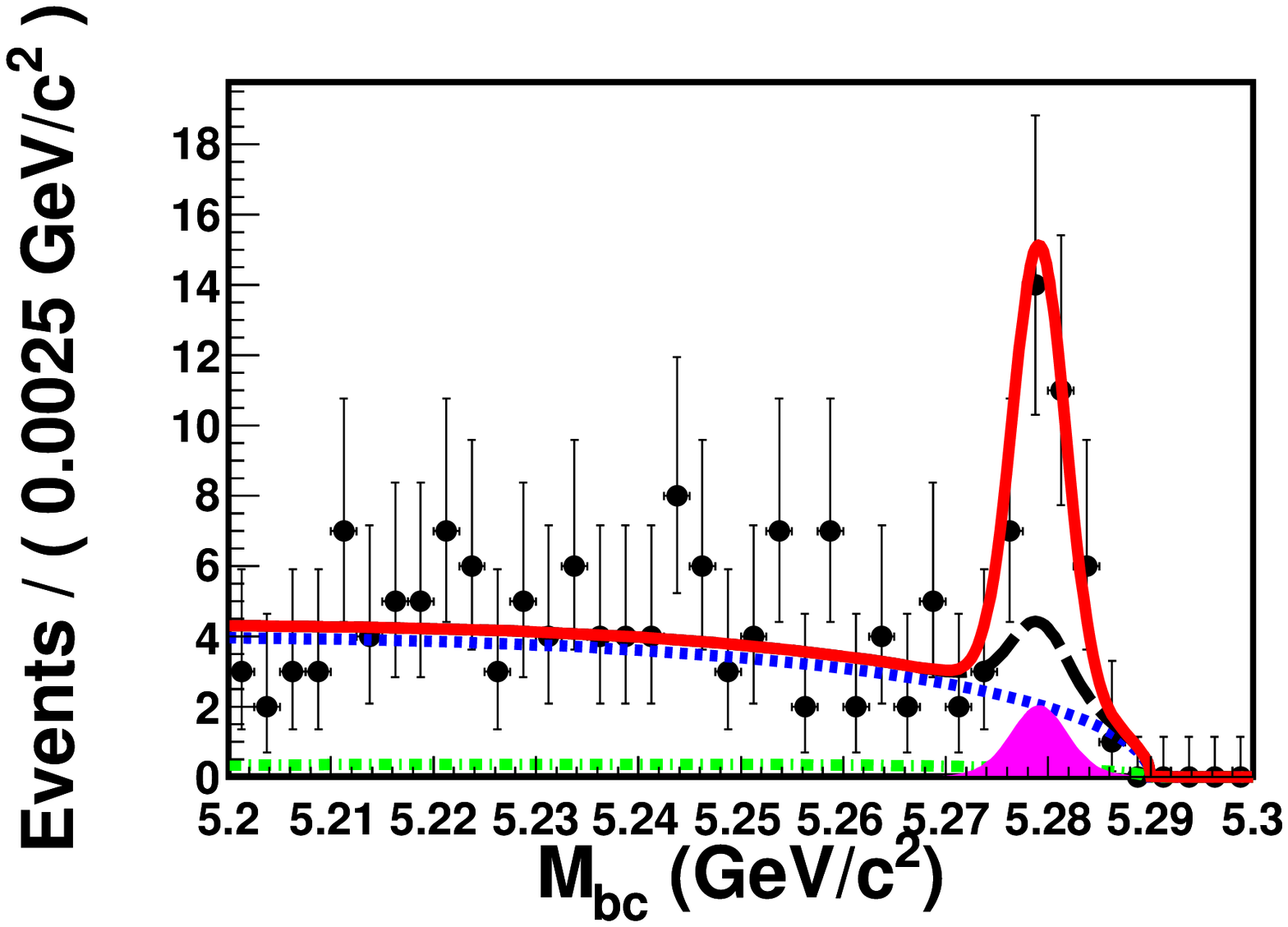}
\end{center}
\caption{The $\Delta E$ and $M_{\rm bc}$ projections 
for $\btopkpg$ (upper) and $\btopksg$ (lower). 
The points with error bars represent the data. The different curves 
show the total fit function (solid red),   
total background function (long-dashed black), 
continuum component (dotted blue), 
the $b\to c$ component (dashed-dotted green) and
the non-resonant component as well as other charmless 
backgrounds (filled magenta histogram).}
\label{fig:dembc}
\end{figure}

 The signal yield is obtained from an extended
unbinned maximum-likelihood fit to the two-dimensional 
$\Delta E$-$M_{\rm bc}$ distribution in the fit region.
We model the shape for the signal component using the product of a 
Crystal Ball line shape~\cite{cbshape} for $\Delta E$ and 
a single Gaussian for $M_{\rm bc}$. 
The continuum background is modeled with a product of 
first order Chebyshev polynomial
for $\Delta E$ and an ARGUS~\cite{argus} function for $M_{\rm bc}$.
The $b \to c$ background is modeled with a product of 
second order Chebyshev polynomial
for $\Delta E$ and an ARGUS plus Gaussian function for $M_{\rm bc}$. 
The small charmless backgrounds (except the non-resonant component) 
are modeled with 
a functional form that is the product of two Gaussians
for $\Delta E$ and with a single Gaussian for
$M_{\rm bc}$~\cite{charmless-pdf}.
In the final fit the continuum parameters are allowed to vary 
while all other
background parameters are fixed to the values from MC simulation. 
The shape of 
the peaking backgrounds are fixed to that of signal in $M_{\rm bc}$
and $\Delta E$.
In the $\btopkpg$ mode, the non-resonant 
background yield is fixed to 
the value from the $\phi$ sideband and assuming isospin symmetry, 
the same non-resonant fraction is used in the neutral mode.
The signal shapes are  adjusted for small differences between MC and data 
using a high statistics 
$B^0 \to K^{*}(892)^0(\to K^+\pi^-) \gamma$ control sample.
The invariant mass of the $K^{*}$ candidates are required to satisfy
0.820 GeV/$c^2 < M_{K^+\pi^-} < 0.970$ GeV/$c^2$. 
The fit yields a signal of 
$136\pm17$ $\btopkpg$ and $35\pm8$ $\btopksg$ candidates.
The projections of the fit results onto $\Delta E$ 
and $M_{\rm bc}$ are shown in Fig.~\ref{fig:dembc}.
The signal significance is defined as 
$\sqrt{-2\,\ln({\cal L}_0/{\cal L}_{\rm max})}$,
where ${\cal L}_{\rm max}$ is
the maximum likelihood for the best fit and ${\cal L}_0$ 
is the corresponding value with the signal yield fixed to zero.
The additive sources of systematic uncertainty described below are
included in the significance by 
varying each by its error and taking the lowest significance.
The signal in the charged mode has a significance of
$9.6\,\sigma$, whereas that for the neutral mode is $5.4\,\sigma$.
\par
\begin{figure}[htbp]
\begin{center}
\includegraphics[width=8.1cm]{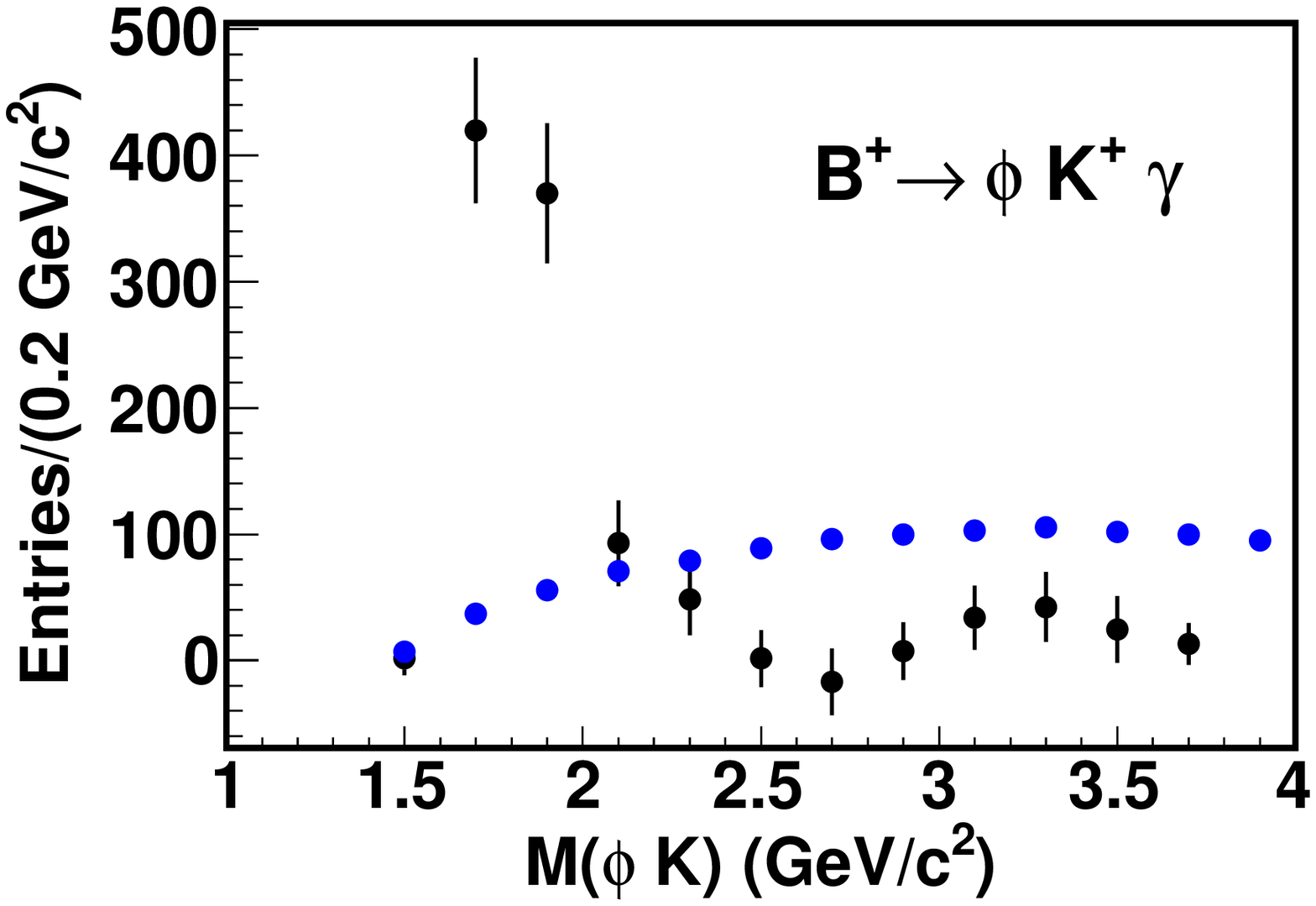}
\includegraphics[width=8.1cm]{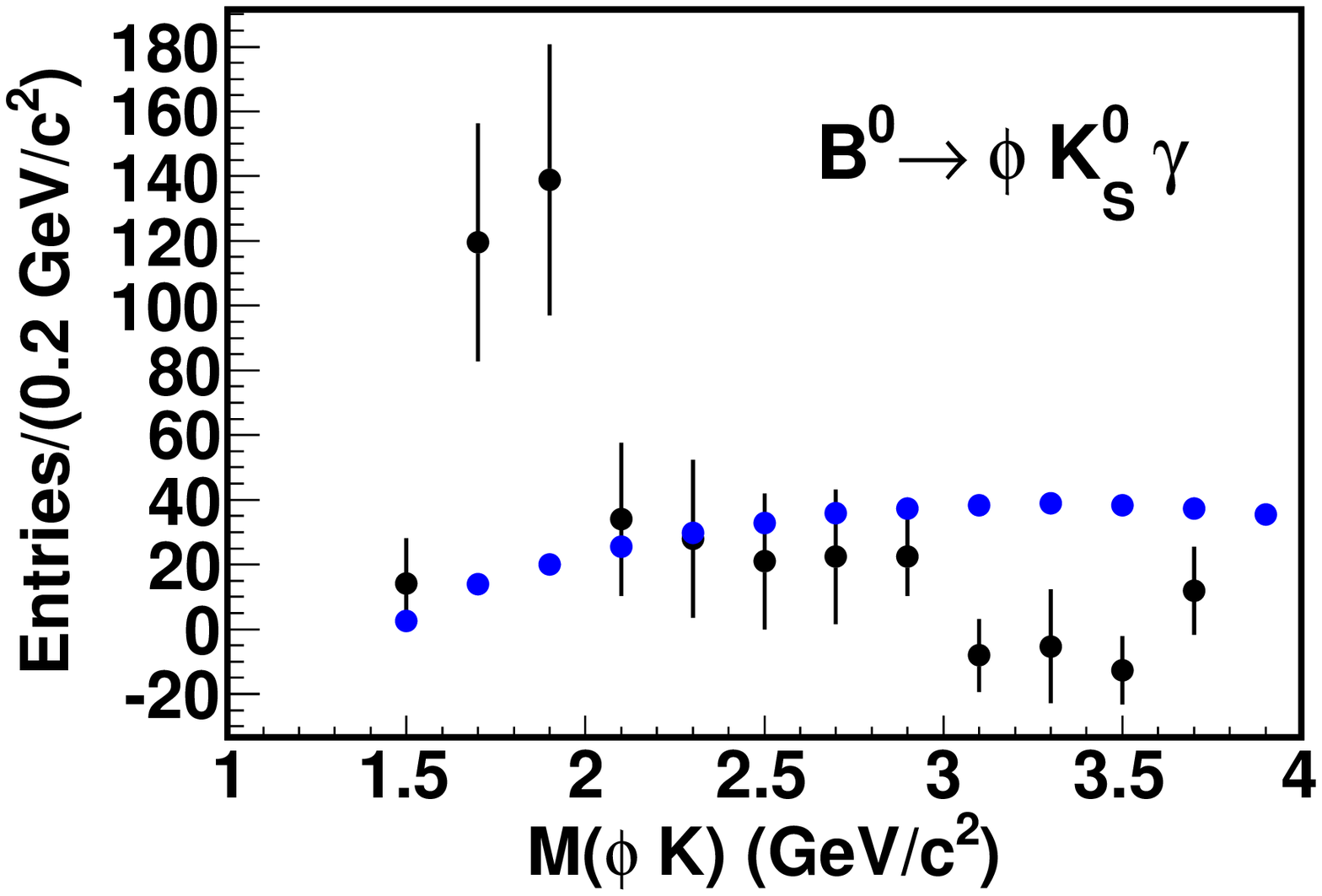}
\end{center}
\caption{The background-subtracted and efficiency-corrected 
$\phi K$ mass distributions for $\btopkpg$ (left) 
and $\btopksg$ (right). The points with error bars 
represent the data. The yield in each bin is obtained by 
the fitting procedure described in the text. 
The three-body phase-space model from the MC simulation 
is shown by the circles (blue) and normalized to the total 
data signal yield.}
\label{fig:mxs}
\end{figure}

 We also examine the $\phi K$ invariant mass distribution
of the signal.
To unfold the $M_{\phi K}$ distribution, 
we subtract all possible backgrounds and correct
the $\phi K$ invariant mass for the efficiency. 
The background-subtracted and efficiency-corrected 
$M_{\phi K}$ distributions are shown in Fig.~\ref{fig:mxs}.
Nearly $72\%$ of the signal events are concentrated
in the low-mass region 
($1.5 \;{\rm GeV}/c^2 < M_{\phi K} <2.0 \;{\rm GeV}/c^2$). 
It is clear that the observed 
$\phi K$ mass spectrum differs significantly from that expected in a 
three-body phase-space decay. The MC-determined reconstruction 
efficiencies 
(defined as the ratio of signal candidates passing all 
selection criteria to the total number of events generated)
are corrected for this $M_{\phi K}$ dependence. 
From the signal yield ($N_{\rm sig}$), we calculate the 
branching fraction ($\mathcal{B}$) as 
$N_{\rm sig}$/
($\epsilon \times N_{\rm B\overline{B}} \times \mathcal{B}_{\rm sec}$), 
where $\epsilon$ is the weighted efficiency, 
$N_{\rm B\overline{B}}$ is the number of $B\overline{B}$ 
pairs in the data sample, and $\mathcal{B}_{\rm sec}$ 
is the product of daughter branching fractions~\cite{pdg}.
The results are summarized in Table~\ref{tab:bfr}.
\begin{table}
\caption{The signal yields, significances, weighted efficiencies and 
branching fractions for the $\btopkpg$ and $\btopkog$ decay modes.}
\begin{ruledtabular}
\begin{tabular}{lcccc}
Decay mode & Yield & Significance ($\sigma$) & Efficiency ($\%$) & Branching fraction ($10^{-6}$) \\
  \hline
$\btopkpg$ & $136\pm17$ & $9.6$  & $15.3\pm0.1$ & $2.34\pm0.29\pm0.23$ \\
$\btopkog$ & $35\pm8$   & $5.4$  & $10.0\pm0.1$ & $2.66\pm0.60\pm0.32$ \\
\end{tabular}
\end{ruledtabular}
\label{tab:bfr}
\end{table}
\par We fit the data with each fixed parameter
varied by its $\pm 1\,\sigma$ error, 
and then the quadratic sum of all differences from the nominal 
value is assigned as the systematic error on the signal yield.
We checked for possible bias in the fitter by doing 
ensemble tests with MC peseudo-experiments.
The statistical errors obtained from our measurements 
are within the expectations from the ensemble tests and 
a systematic error of $0.2\%$ ($2.7\%$) is assigned in 
the $\btopkpg$ ($\btopkog$) mode. The largest contribution comes 
from the non-resonant yield ($8.0\%$).
The total systematic uncertainty assigned to the estimated yield is
$8.2\%$ ($8.8\%$).
We also assign a systematic error of $3.3\%$ ($4.6\%$) 
due to uncertainty on charged track efficiency, 
$1.4\%$ due to particle identification, 
$2.4\%$ due to photon detection efficiency, 
$1.4\%$ due to uncertainty in the number of $B\overline{B}$ pairs in
$\btopkpg$ ($\btopkog$) mode.
Furthermore, we assign a systematic error of $4.6\%$ 
in the neutral mode due to $K_S^0$ reconstruction.
The statistical uncertainty on the MC efficiency 
after reweighting is $0.9\%$ ($1.3\%$). 
The uncertainties due to daughter branching fractions 
account for a systematic contribution of $1.2\%$.
We add each contribution above in quadrature to obtain 
the total systematic uncertainty of $9.9\%$ ($11.9\%$).
\par In summary, we report the first observation of radiative 
$\btopkog$ decays in Belle using a data sample of 
$772 \times 10^6$ $B\overline{B}$ pairs. The observed signal yield is
$35\pm8$ with a significance of $5.4\,\sigma$ 
including systematic uncertainties. The measured branching fraction is
${\cal B}(\btopkog) = (2.66\pm 0.60 \pm 0.32) \times 10^{-6}$.
We also precisely measure
${\cal B}(\btopkpg) = (2.34\pm 0.29 \pm 0.23) \times 10^{-6}$
with a significance of $9.6\,\sigma$. 
The signal events are mostly concentrated at low $\phi K$ mass,
which is similar to a two-body radiative decay.
The neutral mode has enough statistics to measure 
time-dependent $CP$ asymmetry in order to search for new physics 
from right-handed currents in radiative $B$ decays.
\par We thank the KEKB group for the excellent operation of the
accelerator, the KEK cryogenics group for the efficient
operation of the solenoid, and the KEK computer group and
the National Institute of Informatics for valuable computing
and SINET3 network support.  We acknowledge support from
the Ministry of Education, Culture, Sports, Science, and
Technology (MEXT) of Japan, the Japan Society for the 
Promotion of Science (JSPS), and the Tau-Lepton Physics 
Research Center of Nagoya University; 
the Australian Research Council and the Australian 
Department of Industry, Innovation, Science and Research;
the National Natural Science Foundation of China under
contract No.~10575109, 10775142, 10875115 and 10825524; 
the Department of Science and Technology of India; 
the BK21 and WCU program of the Ministry Education Science and
Technology, the CHEP SRC program and Basic Research program (grant No.
R01-2008-000-10477-0) of the Korea Science and Engineering Foundation,
Korea Research Foundation (KRF-2008-313-C00177),
and the Korea Institute of Science and Technology Information;
the Polish Ministry of Science and Higher Education;
the Ministry of Education and Science of the Russian
Federation and the Russian Federal Agency for Atomic Energy;
the Slovenian Research Agency;  the Swiss
National Science Foundation; the National Science Council
and the Ministry of Education of Taiwan; and the U.S.\
Department of Energy.
This work is supported by a Grant-in-Aid from MEXT for 
Science Research in a Priority Area ("New Development of 
Flavor Physics"), and from JSPS for Creative Scientific 
Research ("Evolution of Tau-lepton Physics").

\end{document}

%% file: author-conf2009.tex
\affiliation{Budker Institute of Nuclear Physics, Novosibirsk}
\affiliation{Chiba University, Chiba}
\affiliation{University of Cincinnati, Cincinnati, Ohio 45221}
\affiliation{Department of Physics, Fu Jen Catholic University, Taipei}
\affiliation{Justus-Liebig-Universit\"at Gie\ss{}en, Gie\ss{}en}
\affiliation{The Graduate University for Advanced Studies, Hayama}
\affiliation{Gyeongsang National University, Chinju}
\affiliation{Hanyang University, Seoul}
\affiliation{University of Hawaii, Honolulu, Hawaii 96822}
\affiliation{High Energy Accelerator Research Organization (KEK), Tsukuba}
\affiliation{Hiroshima Institute of Technology, Hiroshima}
\affiliation{University of Illinois at Urbana-Champaign, Urbana, Illinois 61801}
\affiliation{India Institute of Technology Guwahati, Guwahati}
\affiliation{Institute of High Energy Physics, Chinese Academy of Sciences, Beijing}
\affiliation{Institute of High Energy Physics, Vienna}
\affiliation{Institute of High Energy Physics, Protvino}
\affiliation{Institute of Mathematical Sciences, Chennai}
\affiliation{INFN - Sezione di Torino, Torino}
\affiliation{Institute for Theoretical and Experimental Physics, Moscow}
\affiliation{J. Stefan Institute, Ljubljana}
\affiliation{Kanagawa University, Yokohama}
\affiliation{Institut f\"ur Experimentelle Kernphysik, Universit\"at Karlsruhe, Karlsruhe}
\affiliation{Korea University, Seoul}
\affiliation{Kyoto University, Kyoto}
\affiliation{Kyungpook National University, Taegu}
\affiliation{\'Ecole Polytechnique F\'ed\'erale de Lausanne (EPFL), Lausanne}
\affiliation{Faculty of Mathematics and Physics, University of Ljubljana, Ljubljana}
\affiliation{University of Maribor, Maribor}
\affiliation{Max-Planck-Institut f\"ur Physik, M\"unchen}
\affiliation{University of Melbourne, School of Physics, Victoria 3010}
\affiliation{Nagoya University, Nagoya}
\affiliation{Nara University of Education, Nara}
\affiliation{Nara Women's University, Nara}
\affiliation{National Central University, Chung-li}
\affiliation{National United University, Miao Li}
\affiliation{Department of Physics, National Taiwan University, Taipei}
\affiliation{H. Niewodniczanski Institute of Nuclear Physics, Krakow}
\affiliation{Nippon Dental University, Niigata}
\affiliation{Niigata University, Niigata}
\affiliation{University of Nova Gorica, Nova Gorica}
\affiliation{Novosibirsk State University, Novosibirsk}
\affiliation{Osaka City University, Osaka}
\affiliation{Osaka University, Osaka}
\affiliation{Panjab University, Chandigarh}
\affiliation{Peking University, Beijing}
\affiliation{Princeton University, Princeton, New Jersey 08544}
\affiliation{RIKEN BNL Research Center, Upton, New York 11973}
\affiliation{Saga University, Saga}
\affiliation{University of Science and Technology of China, Hefei}
\affiliation{Seoul National University, Seoul}
\affiliation{Shinshu University, Nagano}
\affiliation{Sungkyunkwan University, Suwon}
\affiliation{School of Physics, University of Sydney, NSW 2006}
\affiliation{Tata Institute of Fundamental Research, Mumbai}
\affiliation{Excellence Cluster Universe, Technische Universit\"at M\"unchen, Garching}
\affiliation{Toho University, Funabashi}
\affiliation{Tohoku Gakuin University, Tagajo}
\affiliation{Tohoku University, Sendai}
\affiliation{Department of Physics, University of Tokyo, Tokyo}
\affiliation{Tokyo Institute of Technology, Tokyo}
\affiliation{Tokyo Metropolitan University, Tokyo}
\affiliation{Tokyo University of Agriculture and Technology, Tokyo}
\affiliation{Toyama National College of Maritime Technology, Toyama}
\affiliation{IPNAS, Virginia Polytechnic Institute and State University, Blacksburg, Virginia 24061}
\affiliation{Yonsei University, Seoul}
  \author{I.~Adachi}\affiliation{High Energy Accelerator Research Organization (KEK), Tsukuba} 
  \author{H.~Aihara}\affiliation{Department of Physics, University of Tokyo, Tokyo} 
  \author{K.~Arinstein}\affiliation{Budker Institute of Nuclear Physics, Novosibirsk}\affiliation{Novosibirsk State University, Novosibirsk} 
  \author{T.~Aso}\affiliation{Toyama National College of Maritime Technology, Toyama} 
  \author{V.~Aulchenko}\affiliation{Budker Institute of Nuclear Physics, Novosibirsk}\affiliation{Novosibirsk State University, Novosibirsk} 
  \author{T.~Aushev}\affiliation{\'Ecole Polytechnique F\'ed\'erale de Lausanne (EPFL), Lausanne}\affiliation{Institute for Theoretical and Experimental Physics, Moscow} 
  \author{T.~Aziz}\affiliation{Tata Institute of Fundamental Research, Mumbai} 
  \author{S.~Bahinipati}\affiliation{University of Cincinnati, Cincinnati, Ohio 45221} 
  \author{A.~M.~Bakich}\affiliation{School of Physics, University of Sydney, NSW 2006} 
  \author{V.~Balagura}\affiliation{Institute for Theoretical and Experimental Physics, Moscow} 
% ...
\collaboration{The Belle Collaboration}

%% file: phikgamma.bbl
\begin{thebibliography}{99}
\bibitem{hfag}
E.~Barberio {\it et al.}, Heavy Flavor Averaging Group (HFAG), 
arXiv:0808.1297.
\bibitem{theory_misiak}
M.~Misiak {\it et al.}, Phys. Rev. Lett. {\bf 98}, 022002 (2007).
\bibitem{conj}
Throughout this paper, the inclusion of the charge-conjugate 
decay mode is implied unless otherwise stated.
\bibitem{ags1}
D.~Atwood, M.~Gronau and A.~Soni, Phys. Rev. Lett. {\bf 79}, 185 (1997).
\bibitem{ags2}
D.~Atwood, T.~Gershon, M.~Hazumi and A.~Soni, Phys. Rev. D {\bf 71}, 
076003 (2005).
\bibitem{pol1}
V.~D.~Orlovsky, V.~I. Shevchenko, Phys. Rev. D {\bf 77}, 093003 (2008).
\bibitem{pol2}
D.~Atwood, T.~Gershon, M.~Hazumi and A.~Soni, hep-ph/0701021.
\bibitem{alex_prl}
A.~Drutskoy {\it et al.} (Belle Collaboration), 
Phys. Rev. Lett. {\bf 92}, 051801 (2004).
\bibitem{phikgamma_babar}
B.~Aubert {\it et al.} (BaBar Collaboration), Phys. Rev. D {\bf 75}, 
051102 (2007).
\bibitem{kekb}
S.~Kurokawa and E.~Kikutani, Nucl. Instrum. Methods Phys. Res., Sect. A 
{\bf 499}, 1 (2003), and other papers included in this volume.
\bibitem{Belle}
A.~Abashian {\it et al.} (Belle Collaboration), 
Nucl. Instrum. Methods Phys. Res., Sect. A {\bf 479}, 117 (2002).
\bibitem{evtgen} 
We use the EvtGen $B$ meson decay generator, D.~J.~Lange, 
Nucl. Instrum. Methods Phys. Res., Sect. A {\bf 462}, 152 (2001).
The detector response is simulated with GEANT, R.~Brun {\it et al.}, 
GEANT 3.21, CERN Report DD/EE/84-1 (1984).
\bibitem{pdg}
C.~Amsler, {\it et al.}, Physics Letters {\bf B 667}, 1 (2008).
\bibitem{belle_b2s}
K-F.~Chen {\it et al.} (Belle Collaboration), Phys. Rev. D {\bf 72}, 
012004 (2005).
\bibitem{pi0etaveto}
P.~Koppenburg {\it et al.} (Belle Collaboration), 
Phys. Rev. Lett. {\bf 93}, 061803 (2004).
\bibitem{fisher} 
R.~A.~Fisher, Annals of Eugenics {\bf 7}, 179 (1936).
\bibitem{fox} 
The Fox-Wolfram moments were introduced in
G.~C.~Fox and S.~Wolfram, Phys. Rev. Lett. {\bf 41}, 1581 (1978).
The modifiex Fox-Wolfram moments used in this analysis are described in
S.~H.~Lee {\it et al.} (Belle Collaboration), Phys. Rev. Lett. {\bf 91}, 
261801 (2003).
\bibitem{cbshape} 
T. Skwarnicki, Ph.D. thesis, Institute for Nuclear Physics,
Krakow, 1986; DESY Internal Report No. DESY F31-86-02, 1986. 
The function is widely used to describe asymmetric
distributions caused by shower leakage in crystal calorimeters.
\bibitem{argus} 
H.~Albrecht {\it et al.} (ARGUS Collaboration), Phys. Lett. B {\bf 241}, 
278 (1990).
\bibitem{charmless-pdf}
$
f(\Delta E, M_{\rm bc}) = G_1(\Delta E - E_1) \,G(M_{\rm bc}-M_0) +
G_2(\Delta E - E_2) \,G(M_{\rm bc}-M_0)
$
, where $G_1$, $G_2$ and $G_3$ are Gaussian functions and 
$E_1$, $E_2$ and $M_0$ are constants.

\end{thebibliography}
